\documentclass{article}
\usepackage{spconf,amsmath,graphicx,hyperref,booktabs,multirow,xcolor,enumitem,tikz}
\usetikzlibrary{positioning,arrows.meta,calc,backgrounds}
\usepackage{fontawesome5,pifont}
\hypersetup{colorlinks=true,allcolors=blue!50!black} %
\IfFileExists{figures/colors.tex}{\definecolor{ours}{HTML}{D55E00}
\definecolor{teacher}{HTML}{EAAE7F}
\definecolor{baseline}{HTML}{6B8FB8}
\definecolor{snrours}{HTML}{D55E00}
\definecolor{snresp}{HTML}{0072B2}
\definecolor{snrnoisereduce}{HTML}{555555}
\definecolor{snraudiosep}{HTML}{009E73}
\definecolor{snrtext}{HTML}{E69F00}
\definecolor{snroracle}{HTML}{999999}
\definecolor{snrteacher}{HTML}{CC79A7}
\definecolor{snrnrvad}{HTML}{56B4E9}
}{}
\IfFileExists{figures/overview/colors.tex}{\definecolor{figspan}{HTML}{2B8CD6}
\definecolor{figthreshold}{HTML}{D9541A}
\definecolor{figfail}{HTML}{E03131}
\definecolor{figworks}{HTML}{2F9E44}
}{}

\renewcommand\paragraph[1]{\par\smallskip\noindent\textbf{#1}\ }

\newif\iftodos
\todosfalse

\newcommand{\pending}[1]{\textcolor{blue}{[PENDING: #1]}}

\newcommand{\IfTable}[1]{\IfFileExists{tables/#1.tex}{\input{tables/#1}}{\pending{\detokenize{tables/#1.tex}}}}
\newcommand{\IfFigure}[2]{\IfFileExists{figures/#1}{\includegraphics[width=#2]{figures/#1}}{\fbox{\parbox{#2}{\centering\pending{\detokenize{figures/#1}}\vspace{3em}}}}}
\IfFileExists{tables/numbers.tex}{\newcommand{\NumClips}{47}

\newcommand{\NumMixtures}{188}

}{}
\providecommand{\NumClips}{62}

\newcommand{\ModelName}{CallCleaner}
\newcommand{\SAMvad}{SAM-Audio$_\text{VAD}$}    %
\newcommand{\SAMtext}{SAM-Audio$_\text{text}$}  %
\newcommand{\NRvad}{Noisereduce$_\text{VAD}$}   %
\newcommand{\NRvads}{Noisereduce$^\text{s}_\text{VAD}$} %
\newcommand{\Added}{\textsc{Added}}
\newcommand{\Missing}{\textsc{Missing}}
\newcommand{\Shape}{\textsc{Shape}}
\newcommand{\Outside}{\textsc{Outside}}

\title{CallCleaner: Lightweight Bioacoustic Denoising from an Audio Foundation Model}

\newif\ifanon
\anonfalse
\newcommand{\ProjectURL}{https://callcleaner.github.io}
\ifanon
  \name{Anonymous submission}
  \address{}
\else
  \name{Michael Xie$^{1}$, Ga\v{s}per Begu\v{s}$^{1,2}$, Morris Alper$^{3}$}
  \address{\small
  $^{1}$University of California, Berkeley, USA
  \;\; $^{2}$Project CETI, USA and Dominica
  \;\; $^{3}$University of Miami, USA
  \\[0.4em] \normalsize\url{\ProjectURL}
  }
\fi

\begin{document}
\maketitle

\begin{abstract}
Many studies of animal vocalizations require clean recordings, but animals are usually recorded in the field, with background noise from weather, vehicles, and other animals. \emph{Animal vocalization denoising} is challenged by the scarcity of clean training data and the extreme diversity of animal vocalizations.
Classical spectral gates leave residual noise, especially without voice activity detection (VAD), while detector-free neural denoisers designed for bioacoustics remove less noise.
We show that a general-purpose source separation foundation model, SAM-Audio, removes more noise than either of these types of methods, when used without training and prompted with time spans from a simple animal VAD.
To turn this heavy method into a practical tool, we distill it into \emph{\ModelName}, a 19M-parameter denoiser that needs no detector and runs ${\sim}90\times$ faster than real time on a laptop CPU.
We find that SI-SDR has limited utility in our setting as it penalizes regenerated output and conflates noise removal with call fidelity, a trade-off our separate metrics reveal.
\ModelName{} leaves 11\,dB less noise around the call than Biodenoising, the leading neural bioacoustic denoiser with public weights, and among fast, CPU-capable systems it pushes the Pareto frontier of noise removal vs.\ call fidelity, without requiring a detector.
\ifanon We will publicly release \else We release \fi code and weights for \ModelName{}, which enables fast, private, local denoising of single-caller recordings without specialized hardware.
\end{abstract}


\section{Introduction}
\label{sec:intro}

The animal kingdom exhibits a wide array of vocalizations, used to attract mates, defend territory, warn of predators and coordinate with their group~\cite{bradbury2011principles}. Many uses of these calls require clean audio, for instance as stimuli in playback experiments~\cite{seyfarth1980monkey}. A key challenge is posed by acoustic conditions in the field: background wind and rain, human-created noise such as vehicles and ship engines, and other animals in the background (birds, insects) are inevitable. Since clean studio recording conditions are not usually achievable, the task of \emph{animal vocalization denoising}~\cite{miron2025biodenoising} aims to remove this noise.

This task differs from human speech enhancement in two ways: animal calls cover a diverse range of frequencies and structures, and we lack abundant, clean data as we have for speech. Existing animal vocalization denoisers therefore either use training-free classical spectral gates~\cite{tim_sainburg_2019_3243139,sainburg2020finding}, or are trained on noisy or pseudo-clean targets~\cite{bergler2020orcaclean,miron2025biodenoising,song2026biosen,abbasi2026mice}, and in practice they leave residual noise in their output.

\emph{Foundation models}~\cite{bommasani2021opportunities} now exist for general audio source separation: SAM-Audio~\cite{shi2025samaudio} separates arbitrary sounds in mixed audio, guided by a text prompt or, unlike earlier text-queried separators~\cite{liu2024audiosep,ma2024clapsep}, by time spans. Applying it to bioacoustics is not straightforward: text prompts such as ``animal sound'' are too coarse to single out the call, spans require knowing when the animal vocalizes, and standard metrics misjudge its regenerated output. We ask three questions:

\begin{enumerate}[label=\textbf{Q\arabic*:},nosep,leftmargin=*]
  \item Can foundation source separators denoise animal vocalizations zero-shot?
  \item Can a foundation source separator teach a small, detector-free denoiser that researchers can run locally, without GPUs or sharing sensitive data, and how reliable is it?
  \item How should bioacoustic denoising be evaluated, when systems trade noise removal against call fidelity and some regenerate their output?
\end{enumerate}

\begin{figure*}[t]
  \centering
  \makebox[0.42\textwidth][l]{\small\textbf{(a)}}\hfill\makebox[0.56\textwidth][l]{\small\textbf{(b)}}\\[-2pt]
  \begin{minipage}[c]{0.42\textwidth}
    \includegraphics[width=\linewidth]{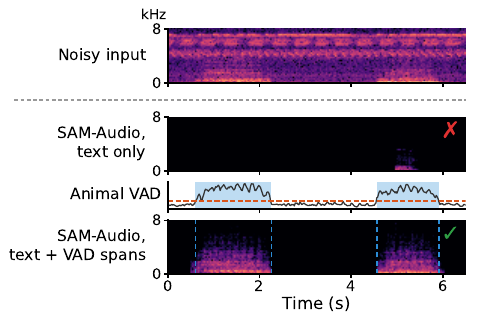}
  \end{minipage}\hfill
  \begin{minipage}[c]{0.56\textwidth}
    \resizebox{\linewidth}{!}{%
    \begin{tikzpicture}[
        font=\scriptsize, node distance=4mm,
        thumb/.style={inner sep=0pt, draw=black!40, line width=0.3pt},
        lbl/.style={font=\scriptsize, align=center, inner sep=1pt, text=black},
        model/.style={align=center, inner sep=3pt},
        taper/.style={draw=black!60, line width=0.5pt},
        tool/.style={draw=black!60, rounded corners=2pt, fill=black!4, align=center, inner sep=3pt},
        arr/.style={-{Latex[length=1.5mm]}, black!70}]
      \node[thumb] (audio) {\includegraphics[width=15mm]{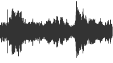}};
      \node[left=1mm of audio, font=\large] (mic) {\faMicrophone};
      \node[thumb, right=14mm of audio] (spans) {\includegraphics[width=15mm]{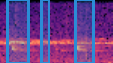}};
      \node[model, minimum height=10mm, right=5mm of spans] (sam) {SAM-Audio\\[-1pt]{\scriptsize (teacher)}};
      \node[thumb, right=5mm of sam] (label) {\includegraphics[width=15mm]{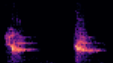}};
      \path (audio.north) ++(0,1.5mm) coordinate (top);
      \foreach \n/\t in {audio/Field audio, spans/Spans, label/Pseudo-labels}
        \node[lbl, anchor=base] at (\n |- top) {\t};
      \draw[arr] (audio) -- node[lbl, above] {Animal\\VAD} (spans);
      \draw[arr] (spans) -- (sam);
      \draw[arr] (sam) -- (label);
      \path (sam.south) ++(0,-11mm) coordinate (row2);
      \coordinate (mid) at ($(audio.east)!0.5!(spans.west)$);
      \node[model, minimum height=8mm] (student) at (mid |- row2) {CallCleaner\\[-1pt]{\scriptsize (student)}};
      \node[font=\huge] (laptop) at (spans |- row2) {\faLaptop};
      \node[lbl, right=1mm of laptop, align=left] {${\sim}38\times$ fewer parameters\\${\sim}90\times$ faster than real time (CPU)};
      \begin{scope}[on background layer]
        \foreach \m/\c in {sam/orange!15, student/green!12}
          \draw[taper, fill=\c] (\m.north west) -- ([yshift=-1.8mm]\m.north east) -- ([yshift=1.8mm]\m.south east)
            -- (\m.south west) -- cycle;
      \end{scope}
      \draw[arr] (audio.south) |- node[lbl, pos=0.75, above] {input} (student.west);
      \draw[arr] (label.south) |- ([yshift=-2.5mm]sam.south) -| node[lbl, pos=0.8, right, xshift=1mm] {target} (student.north);
      \draw[arr] (student) -- (laptop);
    \end{tikzpicture}}
  \end{minipage}
  \caption{\textbf{(a) Zero-shot denoising.} The input (top) is a noisy animal vocalization (here a horse call with two segments). Prompted with ``animal sound'' alone, SAM-Audio misses most of the call (\textcolor{figfail}{\ding{55}}). To specify when the animal vocalizes, an animal VAD marks where its score exceeds a threshold (\textcolor{figthreshold}{orange, dashed}), yielding VAD spans (\textcolor{figspan}{blue}). Prompted with these spans, SAM-Audio recovers the call (\textcolor{figworks}{\ding{51}}; dashed lines: span edges). \textbf{(b) Distillation into CallCleaner.} The zero-shot method in (a) produces pseudo-labels for noisy field audio, used to fine-tune the smaller, faster, and detector-free \ModelName{} student model.}
  \label{fig:overview}
\end{figure*}

For Q1, we prompt SAM-Audio with time spans from an animal voice activity detector (VAD) to specify when the animal vocalizes (Fig.~\ref{fig:overview}a). This shifts much of the problem to detection, which can be performed reasonably with a simple heuristic.
For Q2, we distill SAM-Audio into \emph{\ModelName{}} (Fig.~\ref{fig:overview}b), which matches its teacher's noise removal at Biodenoising's size and speed.
For Q3, we show that SI-SDR has limited utility in our setting, and that important trade-offs are revealed with spectral metrics and separate evaluation of noise removal and call fidelity.
Overall, SAM-Audio removes the most noise of any system, and among systems that run on a laptop CPU, \ModelName{} pushes the Pareto frontier of noise removal vs.\ call fidelity (Fig.~\ref{fig:tradeoff}), enabling fast, private, local enhancement with strong performance at noise removal.

\section{Method}
\label{sec:method}

We target recordings of a single salient foreground animal, whose calls are separated by periods of background noise, as in Biodenoising's task definition~\cite{miron2025biodenoising} (dense and multi-animal calls are discussed as future work in Sec.~\ref{sec:conclusion}). An overview of our method is shown in Fig.~\ref{fig:overview}; we describe its two stages below.

\paragraph{Stage 1: Zero-shot denoising.} SAM-Audio~\cite{shi2025samaudio} is a flow-matching diffusion transformer that generates target and residual streams as audio-codec latents, with optional conditioning signals including text prompts and time spans. When SAM-Audio is prompted with text alone (\SAMtext{}), we find (Sec.~\ref{sec:results}) that it often selects the wrong target; this is largely corrected by also inputting time spans estimated from the noisy input with a simple animal VAD (\emph{max-band}): it uses a 64-band log-mel spectrogram (200\,Hz--8\,kHz) and scores each frame by its 90th-percentile band excess over the estimated noise floor, and keeps frames with robust $z$-score above 1, merging gaps under 0.3\,s. We call this method \SAMvad{}.

\paragraph{Stage 2: Distillation into \ModelName{}.}\label{sec:distill} \SAMvad{} is slow (near real time even on a GPU) and inherits VAD errors. We address these issues by distilling it into a small waveform-to-waveform model, \ModelName{}, that runs on input audio alone (no detector or prompt), initialized from the existing Biodenoising Demucs model~\cite{miron2025biodenoising}. \SAMvad{} serves as the teacher model: for each unlabeled field recording (selected to contain calls with gaps; Sec.~\ref{sec:setup}), its target stream gives a pseudo-label, and its residual stream an estimate of the noise. Half of each batch is remixed by adding another segment's residual to a pseudo-label, as in bootstrapped remixing~\cite{tzinis2022remixit}, which gives training pairs with more varied noise.

\section{Evaluating Denoised Audio}
\label{sec:eval}

Denoisers trade noise removal against call fidelity: a more aggressive system leaves less background but risks removing or distorting parts of the call~\cite{itu2003p835}. However, the standard SI-SDR metric~\cite{leroux2019sdr} compares waveforms sample by sample, penalizing any deviation from the reference and conflating these factors. Such metrics are also known to disfavor generative speech enhancement, since they penalize inaudible changes~\cite{deoliveira2023behavior}. In particular, we find (Sec.~\ref{sec:res-eval}) that SAM-Audio's codec introduces waveform changes which bound achievable SI-SDR even when no separation is performed. Alternative metrics are designed for human speech~\cite{reddy2022dnsmos}, untested on animal vocalizations~\cite{wang2026saj}, or waveform-based rather than spectral~\cite{vincent2006performance}. We therefore adapt log-mel distances~\cite{gray1976distance,kubichek1993mel} and directly measure three separate error sources: noise left around the call (by energy), noise left under the call, and distortion of the call itself (by log-mel distance).

Let $x$, $s$ and $\hat{s}$ be the noisy input, the clean reference and the output. The ground-truth call spans are the 50\,ms frames of $s$ within 20\,dB of its peak. Let $L_z[f,t]$ be the energy of signal $z$, scaled to unit RMS, in mel band $f$ (64 bands, 0--8\,kHz) at frame $t$, in dB, averaged over three STFT resolutions.
With $D = L_{\hat{s}} - L_s$ and $\langle\cdot\rangle$ the mean over bands and in-call frames, we define the log-mel distance inside the call as $\text{LMD}_\text{call} = \langle |D| \rangle$, and its parts as
\begin{equation}
\begin{gathered}
D^{\pm} = \max(\pm D, 0), \qquad
\text{\Shape} = \langle |D - \langle D \rangle| \rangle,\\
\text{\Added} = \langle D^{+} \rangle, \qquad
\text{\Missing} = \langle D^{-} \rangle.
\end{gathered}
\label{eq:parts}
\end{equation}
\Added{} is the noise left under the call, \Missing{} is the call energy the output lacks, and \Shape{} is the error after removing the mean level offset\footnote{\Added{} and \Missing{} normalize to unit RMS over the call region only, so noise left around the call cannot scale the call; LMD$_\text{call}$ normalizes over the whole clip. They therefore do not sum exactly to LMD$_\text{call}$.}. In addition, with $E_z = \sum_{n\notin\mathcal{S}} z[n]^2$ the energy of signal $z$ outside the call span samples $\mathcal{S}$, the noise removed around the call is \Outside{} $= 10\log_{10}(E_{\hat{s}}/E_x)$. It is undefined if the spans cover more than 95\% of the clip.

We define an output as a \emph{catastrophic failure} if $\text{LMD}_\text{call}(\hat{s}) > \text{LMD}_\text{call}(x) + 5$\,dB.

As these are intrusive metrics, they are limited by the quality of the reference. In particular, if the reference contains background noise (as in most field recordings), when a denoiser removes this noise it counts towards \Missing{}, underestimating its performance.

\section{Experimental Setup}
\label{sec:setup}

\paragraph{Data.}
For distillation, we use 4\,s segments from the public corpora Biodenoising~\cite{miron2025biodenoising} was trained on (DCASE few-shot bioacoustics~\cite{nolasco2023fewshot}, Orchive and five bird corpora), none shared with its benchmark, keeping segments with 10--80\% of frames active according to the VAD (1.8\,h in total).

For evaluation, we use Biodenoising's test benchmark: 62 manually selected clean vocalizations from diverse taxa, mixed with real field noise at $-5$, $0$, $5$ and $10$\,dB SNR, at 16\,kHz; we resample all outputs to 16\,kHz for scoring. Main results use the \NumClips{} clips of at most 30\,s, SAM-Audio's training length.

As the training set does not include clean references needed to select settings (VAD, prompt, student recipe, gate smoothing), we select these on the benchmark while reporting robustness to these choices. In particular, 10 of 11 student variants we trained remove more noise than Biodenoising on at least 40 of 47 clips, and selecting the recipe on half the clips and testing on the other half gives the same result (19/21 and 25/26 held-out clips).
SAM-Audio's training data are not publicly listed, and 18 of the \NumClips{} clean calls are in FSD50K~\cite{fonseca2021fsd50k}, so we cannot rule out that SAM-Audio saw these clips during training. However, \ModelName{}'s training pseudo-labels came from separate field recordings, and it still removes more noise than Biodenoising on 26 of the 29 clips not in FSD50K. Overall, there is a need for more clean, openly-available bioacoustic data to provide an independent test set; we discuss this further in Sec.~\ref{sec:conclusion}.

\paragraph{Systems.} For our method, we use \texttt{sam-audio-small} (0.71B parameters; size chosen due to compute limits), reporting a single run of its stochastic sampling (median LMD$_\text{call}$ varies by $\pm0.3$\,dB across three sampling runs). \ModelName{} is fine-tuned from Biodenoising's Demucs model (18.9M parameters) for 2k steps (batch size 16, 1\,s crops, residuals remixed at $-5$ to $10$\,dB SNR); labeling and training take under two hours (Apple M4 Pro, integrated GPU). We train three seeds and report their mean on the benchmark.

We compare to the spectral gate Noisereduce~\cite{tim_sainburg_2019_3243139,sainburg2020finding} in three variants: with its default settings; with a noise profile taken from outside the VAD spans (\NRvad), giving the gate the same timing information as SAM-Audio; and with this noise profile and smoothed masks (\NRvads; 200\,ms, 1\,kHz smoothing replacing the default 50\,ms, 500\,Hz, to reduce pumping artifacts~\cite{berouti1979enhancement,cappe1994elimination}). We also compare to two neural methods, the denoiser Biodenoising~\cite{miron2025biodenoising} and the text-queried separator AudioSep~\cite{liu2024audiosep} (prompted with ``animal sound''). We omit the recent BioSEN~\cite{song2026biosen}, which has not released code or weights and could not be evaluated with our metrics. As an ablation of the teacher, we train the same student on labels from Biodenoising and from \NRvads{} instead of SAM-Audio. We report medians over clips and SNRs and paired Wilcoxon tests over clips (averaging each clip's four mixtures; $p_\text{Holm}$: Holm-corrected within families).

\section{Results}
\label{sec:results}

\begin{figure}[t]
  \centering
  \IfFigure{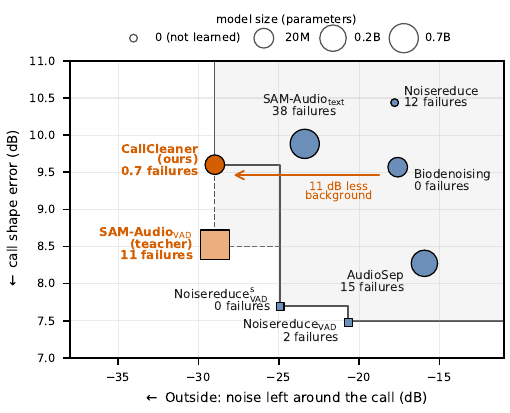}{\linewidth}
  \caption{Benchmark: noise left around the call vs.\ error in the call's shape; lower is better on both axes. Marker area grows with model size. Dashed: Pareto frontier over all systems (shaded: the region it dominates); solid: over systems that run faster than real time on a laptop CPU. Failures: catastrophic failures of \NumMixtures{} mixtures (\ModelName: mean over 3 seeds). Squares: systems that need VAD spans at inference; circles: detector-free. Colors: \textcolor{ours}{\ding{110}}\,\ModelName{}; \textcolor{teacher}{\ding{110}}\,its teacher; \textcolor{baseline}{\ding{110}}\,baselines.}
  \label{fig:tradeoff}
\end{figure}

Our main results, comparing all systems, are shown in Table \ref{tab:main} and Figure \ref{fig:tradeoff}. We proceed to analyze these, addressing our research questions Q1--3 (Sec. \ref{sec:intro}).

\begin{table}[t]
  \centering
  \caption{Benchmark: median scores (dB) over \NumClips{} clips $\times$ 4 SNRs. \textbf{Bold}/\underline{underlined}: best/second best. Metrics are defined in Sec. \ref{sec:eval}. Fail: catastrophic failures of \NumMixtures{} mixtures.}
  \label{tab:main}
  \footnotesize
  \setlength{\tabcolsep}{0.4pt}
  \begin{tabular*}{\linewidth}{@{\extracolsep{\fill}}lccccccc@{}}
\toprule
System & \begin{tabular}[b]{@{}c@{}}SI-SDR\\$\uparrow$\end{tabular} & \begin{tabular}[b]{@{}c@{}}LMD$_\text{call}$\\$\downarrow$\end{tabular} & \begin{tabular}[b]{@{}c@{}}\Outside{}\\$\downarrow$\end{tabular} & \begin{tabular}[b]{@{}c@{}}\Added{}\\$\downarrow$\end{tabular} & \begin{tabular}[b]{@{}c@{}}\Missing{}\\$\downarrow$\end{tabular} & \begin{tabular}[b]{@{}c@{}}\Shape{}\\$\downarrow$\end{tabular} & \begin{tabular}[b]{@{}c@{}}Fail\\$\downarrow$\end{tabular} \\
\midrule
Noisy input & 5.0 & 15.3 & 0.0 & 15.0 & 0.2 & 11.7 & -- \\
\midrule
\multicolumn{8}{@{}l}{\textit{Dedicated denoisers}} \\
Noisereduce & 8.3 & 10.7 & $-$17.8 & \underline{4.8} & 3.6 & 10.4 & 12 \\
Noisereduce$_\text{VAD}$ & \underline{12.1} & \textbf{7.4} & $-$20.7 & \textbf{3.5} & 2.8 & \textbf{7.5} & 2 \\
Noisereduce$^\text{s}_\text{VAD}$ & 10.6 & \underline{7.8} & \underline{$-$24.9} & 5.3 & 1.7 & \underline{7.7} & \textbf{0} \\
Biodenoising & \textbf{13.0} & 10.7 & $-$17.6 & 9.7 & \textbf{0.7} & 9.6 & \textbf{0} \\
\midrule
\multicolumn{8}{@{}l}{\textit{Zero-shot separators}} \\
AudioSep, text & 10.3 & 8.9 & $-$15.9 & 6.8 & \underline{1.2} & 8.3 & 15 \\
SAM-Audio$_\text{text}$ & $-$0.3 & 10.9 & $-$23.4 & 6.2 & 2.2 & 9.9 & 38 \\
SAM-Audio$_\text{VAD}$ & 1.4 & 8.8 & \textbf{$-$29.0} & 5.1 & 2.1 & 8.5 & 11 \\
\midrule
\multicolumn{8}{@{}l}{\textit{Distilled students}} \\
CallCleaner (ours) & \textbf{13.0} & 9.6 & \textbf{$-$29.0} & 7.2 & 1.9 & 9.6 & \underline{0.7} \\
\bottomrule
\end{tabular*}

\end{table}

\subsection{Zero-shot separation (Q1)}

Q1 asked whether foundation source separation can denoise animal vocalizations zero-shot. SAM-Audio already removes more noise than most dedicated methods from a text prompt alone, as measured by the \Outside{} metric, but it is brittle, returning the wrong stream (its residual is closer to the clean call than its target) for 15--23\% of mixtures (Fig.~\ref{fig:overview}a). This is mitigated by inputting VAD time spans; \SAMvad{} removes the most noise of any system (\Outside{} $-29.0$\,dB) while staying close to the best systems inside the call, behind only the VAD gates in LMD$_\text{call}$ (Table~\ref{tab:main}). Its advantage does not come from the spans alone: masking Biodenoising's output to the same spans removes noise but cuts call frames the VAD misses (LMD$_\text{call}$ 12.5\,dB).

\SAMvad{} still fails catastrophically on 11 mixtures (Table~\ref{tab:main}), and we observe that this often occurs when VAD misses portions of the call. Our simple, training-free detector already fixes most of SAM-Audio's wrong-stream errors (catastrophic failures 38$\rightarrow$11; Table~\ref{tab:main}), but better detection is expected to mitigate these failures.

\subsection{\ModelName{} on the benchmark (Q2)}
\label{sec:res-student}

Q2 asked whether a foundation separator can teach a small, detector-free denoiser (\ModelName{}) that researchers can run locally, and how reliable it is.

Regarding denoising performance, we find that \ModelName{} matches its teacher's noise removal, leaving 11\,dB less noise around the call than Biodenoising (\Outside{}; better on 44/47 clips, $p<10^{-10}$). This holds when normalizing for in-call energy: \ModelName{} still leaves 11.2\,dB less noise around the call than Biodenoising (better on 42/47 clips), so the improvement is not due to overall output level. Students trained with \ModelName{}'s recipe on pseudo-labels from Biodenoising or the gate do not match this noise removal (\Outside{} $-17.6$ and $-23.2$\,dB), so the gain comes from the teacher. Inside the call, \ModelName{} performs similarly to Biodenoising; its small lead (LMD$_\text{call}$ 9.6 vs.\ 10.7\,dB, $p=0.002$; Table~\ref{tab:main}) comes from leaving less noise under the call (\Added{}) outside the call's frequency band, and vanishes when both outputs are band-passed to the call's band (the range holding 95\% of the clean call's energy; 8.5 vs.\ 8.4\,dB).

Regarding local, lightweight use and reliability: \ModelName{} requires no VAD and rarely fails badly (0.7 catastrophic failures per seed, vs.\ 11 for its teacher \SAMvad{}; Table~\ref{tab:main}). Since it is a predictive rather than regenerative model, it rarely shifts the pitch of tonal calls: F0 (pYIN, 37 tonal clips) moves by more than 50\,cents on 3\% of mixtures, as with the gate, vs.\ 22\% for the teacher. Its call shape, however, matches Biodenoising's rather than its teacher's (\Shape{} 9.6 vs.\ 8.5\,dB; Table~\ref{tab:main}). It runs about 90$\times$ faster than real time on a laptop CPU (Apple M4 Pro), as fast as Biodenoising, while \texttt{sam-audio-small} runs only near real time on the same laptop's integrated GPU.

While \ModelName{} shows strong performance, we note a trade-off relative to classical spectral gates (Table~\ref{tab:main}): \NRvads{} keeps the call's shape better (\Shape{}; better on 34/47 clips, $p_\text{Holm}=0.02$) and never fails catastrophically, but it requires the VAD. \ModelName{} removes more noise (\Outside{}; 31/47 clips, $p_\text{Holm}=0.03$), though mostly outside the call's band (band-passed, it leads on only 27/47 clips, $p=0.27$). Among systems that run on a CPU, \ModelName{} and the VAD gates span the frontier (Fig.~\ref{fig:tradeoff}). This matches prior findings~\cite{mcewen2023automatic} that classical gates remain competitive with neural methods, which trade simplicity for additional noise reduction. Which to use in practice depends on whether a reliable detector is available, and on whether the call's shape or the background matters more.

\subsection{Evaluation (Q3)}
\label{sec:res-eval}

Q3 asked how denoising should be evaluated when systems trade noise removal against call fidelity and some, like SAM-Audio, regenerate their output. Our findings confirm that SI-SDR is biased against regeneration. Passing clean calls through SAM-Audio's codec alone, with no separation, yields a median SI-SDR of only 7.2\,dB, although the output is not shifted in time (measured via cross-correlation) and its LMD$_\text{call}$ of 3.3\,dB is better than any system's. Even a perfect SAM-Audio separation would thus rank below Biodenoising (13.0\,dB; Table~\ref{tab:main}) on SI-SDR, so SI-SDR alone would have ruled out SAM-Audio as a teacher; however, our results show that \SAMvad{} serves as a better teacher than Biodenoising (Sec.~\ref{sec:res-student}), and its student \ModelName{}, which computes its output from the input waveform rather than regenerating it, ties Biodenoising for the best SI-SDR. The best teacher was thus the system that removes the most noise, despite SI-SDR ranking it last. SI-SDR also hides the trade-off between noise removal and call fidelity: it rates \ModelName{} level with Biodenoising despite 11\,dB more noise removal. Since calls are sparse, noise removal both between and within calls contributes substantially to SI-SDR. Zeroing the noisy input outside the ground-truth call spans, without removing any noise within them, already improves its SI-SDR by 3.3\,dB (median), a third of Biodenoising's 10.4\,dB improvement. This is another reason to score the two regions separately.

Even spectral in-call scores favor filters over regeneration: against the clean call passed through SAM-Audio's codec, \SAMvad{} overtakes \NRvad{} inside the call (LMD$_\text{call}$ 6.2 vs.\ 8.4\,dB), reversing their ranking against the clean reference. \Outside{}, which compares the output with the input rather than with a reference, avoids this dependence, but it must be read alongside the in-call metrics, since silencing everything would also score well. Ultimately, both noise removal and call fidelity need listening tests.

\section{Conclusion}
\label{sec:conclusion}

A foundation source separator prompted with call spans from a simple detector removes more noise from animal vocalizations than dedicated denoisers. When distilled into our small, detector-free model \ModelName{}, it keeps this noise removal while running much faster than real time on a laptop CPU. Scoring the two separately exposes their trade-off and highlights the different use cases for \ModelName{} and smoothed spectral gates.

\paragraph{Limitations and future work.} We have focused on sparse calls with one salient animal; future work could extend \ModelName{} to dense targets and condition it on which caller to keep. Our heuristic VAD could be improved in line with work on self-supervised audio encodings for bioacoustic event detection~\cite{bermant2022bioacoustic,bharadwaj2025openbeats}. The student may also inherit pseudo-label errors; future work could explore recent mitigation strategies~\cite{cornell2026snoar}. Extremely low- and high-frequency calls and impulsive noise remain challenging. Since our settings were chosen on the benchmark and its recordings may have been seen by SAM-Audio (Section \ref{sec:setup}), larger, open collections of clean animal vocalizations are needed as independent test data. Last, like prior work, our evaluation reflects human hearing, while animals may perceive denoised calls differently~\cite{dooling2002auditory}.

\section{Acknowledgements}
We thank Letizia Cerqueglini, Yakov Kolani, Kevin Wang, and Samuele Cornell for helpful feedback. This project was partially funded by Project CETI and Schmidt Sciences' Humanities and AI Virtual Institute.

\section{Generative AI Use Disclosure}
Generative AI tools were used for writing and coding assistance.

\bibliographystyle{IEEEbib}
\bibliography{refs}

\end{document}